\documentclass{article}

\def\be{\begin{equation}}
\def\ee{\end{equation}}
\def\bea{\begin{eqnarray}}
\def\eea{\end{eqnarray}}
\begin{document}
\begin{titlepage}
\vspace*{10mm}
\begin{center}
{\large \bf Logarithmic two dimensional spin--1/3 fractional supersymmetric
conformal field theories and the two point functions} \vskip 10mm
\centerline {\bf
 Fardin Kheirandish $^{a}$ \footnote {e-mail:fardin@iasbs.ac.ir} {\rm and}
 Mohammad Khorrami $^{a,b}$ \footnote {e-mail:mamwad@iasbs.ac.ir}}
 \vskip 1cm
{\it $^a$ Institute for Advanced Studies in Basic Sciences,}
\\ {\it P. O. Box 159, Zanjan 45195, Iran}\\
{\it $^b$ Institute for Studies in Theoretical physics and
Mathematics,}\\ {\it P. O. Box 5531, Tehran 19395, Iran}\\
\end{center}
\vskip 2cm
\begin{abstract}
\noindent Logarithmic spin--1/3 superconformal field theories are
investigated. The chiral and full two--point functions of two- (or more-)
dimensional Jordanian blocks of arbitrary weights, are obtained.
\end{abstract}
\end{titlepage}
\newpage
\section{Introduction}
According to Gurarie [1], conformal field theories which their
correlation functions exhibit logarithmic behaviour may be
consistently defined. In some interesting physical theories like
polymers [2], WZNW models [3--6], percolation [7], the
Haldane-Rezayi quantum Hall state [8], and edge excitation in
fractional quantum Hall effect [9], logarithmic
correlation functions appear. Also the logarithmic operators can be
considered in 2D-magnetohydrodynamic turbulence [10,11,12],
2D-turbulence ([13], [14]) and some critical disordered models
[15,16]. Logarithmic conformal field theories for D dimensional
case ($D>2$) has also been studied [17]. In this paper we consider
a superconformal extension of Virasoro algebra [18,19] corresponding to
three--component supermultiplets, and then, following [20--22], generalize
the superconformal field to Jordanian blocks of quasi superconformal fields.
We then find the two--point functions of chiral- and full-component fields.
It is seen that this correlators are readily obtained through formal
derivatives of correlators of superprimary fields, just as was seen in
[20--22].
\section{Superprimary and quasi-superprimary fields}
A chiral superprimary field $\Phi(z,\theta)$ with conformal weight
$\Delta$, is an operator satisfying [18]
\be
[L_n, \Phi(z,\theta)]=\left[ z^{n+1}\partial_{z}+(n+1)\left(\Delta+
\frac{\Lambda}{3}\right)z^{n}\right]\Phi(z,\theta),
\ee
\be
[G_{r},
\Phi(z,\theta)]=[z^{r+1/3}\delta_{\theta}-q^{2}z^{r+1/3}\theta^{2}
\partial_{z}-(3r+1)q^{2}\Delta z^{r-2/3}\theta^{2}]\Phi(z,\theta),
\ee
where $\theta$ is a paragrassmann variable, satisfying $\theta^{3}=0$,
$q$ is one of the third roots of unity, not equal to one, and
$\delta_{\theta}$ and $\Lambda$, satisfy [18,19]
\be
\delta_{\theta}\theta=q^{-1}\theta\delta_{\theta}+1,
\ee
and
\be
[\Lambda, \theta]=\theta,\hspace{1
cm}[\Lambda,\delta_{\theta}]=-\delta_{\theta}.
\ee

Here $L_{n}$'s and $G_{r}$'s are the generators of the
supervirasoro algebra satisfying \bea
[L_{n},L_{m}]&=&(n-m)L_{n+m},\cr
[L_{n},G_{r}]&=&\left({n\over 3}-r\right)G_{n+r},
\eea
and
\be
G_rG_sG_t+\hbox{five other permutations of the
indices}=6L_{r+s+t}.
\ee
The superprimary field $\Phi(z,\theta)$, is written as
\be
\Phi:=\Phi(z,\theta)=\varphi(z,\theta)+\theta\varphi_{1}(z)+
\theta^{2}\varphi_{2}(z),
\ee
where $\varphi_{1}(z)$ and $\varphi_{2}(z)$ are paragrassmann fields of
grades 1 and 2, respectively. One can similarly define a
complete superprimary field $\Phi(z, \bar{z}, \theta,
\bar{\theta})$ with the weights $(\Delta, \bar{\Delta})$ and the expansion
\be
\Phi=\sum_{k,
k'=0}^{2}\theta^{k}\bar{\theta}^{k'}\varphi_{kk'},
\ee
through (1) and (2), and obvious analogous relations with
$\bar{L}_{n}$'s and $\bar{G}_{r}$'s. Now suppose that the first
component field $\varphi(z)$ in chiral superprimary field
$\Phi(z,\theta)$, has a logarithmic counterpart $\varphi'(z)$ [20]:
\be
[L_{n}, \varphi'(z)]=[z^{n+1}\partial_{z}+(n+1)z^{n}\Delta]
\varphi'(z)+(n+1)z^{n}\varphi(z).
\ee
We will show that $\varphi'(z)$ is the first component field of a new
superfield $\Phi'(z,\theta)$, which is the formal derivative of the
superfield $\Phi(z,\theta)$ with respect to its weight. Let us define the
fields $f'_{r}(z)$ by
\be
[G_{r}, \varphi'(z)]=:z^{r+1/3}f'_{r}(z),
\ee
where $r+\frac{1}{3}$ is an integer. Following [22], acting on the both
sides of the above equation with $L_{m}$ and using the Jacobi identity, and
using (9), (1), and (5), we have
\bea
[L_{m},f'_{r}(z)]&=&\left({m\over 3}-r\right)z^{m}[f'_{m+r}(z)-f'_{r}(z)]+
\Big[z^{m+1}\partial_{z}\cr
&&+(m+1)\left(\Delta+{1\over 3}\right) z^{m}\Big]f'_{r}(z)+
(m+1)z^{m}\varphi_{1}(z).
\eea
Demanding
\be
[L_{-1}, f'_{r}(z)]=\partial_{z}f'_{r}(z),
\ee
it is easy to shown that
\be
f'_r(z)=\cases{\psi'(z),&$r\geq -1/3$\cr \psi''(z),&$r\leq
-4/3.$\cr}
\ee
Then, equating $[L_{1},f'_{-4/3}(z)]$ and $[L_{1},f'_{-7/3}(z)]$, we obtain
\be
\psi'(z)=\psi''(z)=:\psi'_{1}.
\ee
So in this way we obtain a well--defined field $\psi'_{1}$, satisfying
\be
[G_{r},\varphi']=z^{r+1/3}\psi'_{1},
\ee
\be
[L_{n},\psi'_{1}]=\left[z^{n+1}\partial_{z}+(n+1)z^{n}\left(\Delta +
{1\over 3}\right)\right]\psi'_{1}+(n+1)z^{n}\psi_{1}.
\ee
Again, let's define the fields $h'_{r}(z)$ through
\be
[G_{r},\psi'_{1}]_{q^{-1}}:=-z^{r+1/3}h'_{r}(z).
\ee
Acting both sides
with $L_{m}$ and using the generalized Jacobi identity [18]:
\be
[[G_{r},\psi'_{1}]_{q^{-1}},L_{m}]+[G_r,[L_m,\psi'_{1}]]_{q^{-1}}+
[[L_{m},G_{r}],\psi'_{1}]_{q^{-1}}=0,
\ee
we obtain
\bea
[L_{m},h'_{r}]&=&\left[z^{m+1}\partial_{z}+(m+1)\left(\Delta+{1\over 3}
\right)z^{m}\right]h'_{r}+(m+1)z^{m}\psi_{2}\cr
&&+\left({m\over 3}-r\right)z^{m}h'_{m+r}+\left(r+{1\over 3}\right)
z^{m}h'_{r}.
\eea
Then, using the same method applied to determine the form of the functions
$f'_{r}(z)$, we find a well--defined field $\psi'_{2}$ satisfying
\be
[G_{r},\psi'_{1}]_{q^{-1}}=-z^{r+1/3}\psi'_{2},
\ee
\be
[L_{n},\psi'_{2}]=\left[z^{n+1}\partial_{z}+(n+1)\left(\Delta +{2\over 3}
\right)z^{n}\right]\psi'_{2}+(n+1)z^{n}\psi_{2}.
\ee
Finally, we must calculate $[G_{r},\psi'_{2}]_{q^{-2}}$. Substituting
for $\psi'_{2}(z)$ from (20) and (15), and using (6), we have
\bea
[G_{r},\psi'_{2}]_{q}&=&-[z^{r+1/3}\partial_{z}\varphi'(z)+(3r+1)z^{r-2/3}
\Delta\varphi'(z)\cr &&+(3r+1)z^{r-2/3}\varphi(z)].
\eea
Now we define the quasi superprimary field $\Phi'$:
\be
\Phi':=\Phi'(z,\theta)=\varphi'(z)+\theta\psi'_{1}(z)+\theta^{2}\psi'_2(z).
\ee
It is easy to see that
\be
[L_{n},\Phi']=\left[z^{n+1}\partial_{z}+(n+1)z^{n}\left(\Delta+
\frac{\Lambda}{3}\right)\right]\Phi'+(n+1)z^{n}\Phi,
\ee
\be
[G_{r},\Phi']=[z^{r+1/3}(\delta_{\theta}-q^{2}\theta^{2}\partial_{z})-
(3r+1)z^{r-2/3}q^{2}\Delta\theta^{2}]\Phi'
-q^{2}(3r+1)z^{r-2/3}\theta^{2}\Phi.
\ee
We see that (24) and (25) are formal derivatives of (1) and (2) with respect
to $\Delta$, provided one defines the formal derivative [20--22]
\be
\Phi'(z,\theta) =:\frac{d\Phi}{d\Delta}.
\ee
The two superfields $\Phi$ and $\Phi'$, are a two dimensional Jordanian
block of quasi-primary fields. The generalization of the above results
to an $m$ dimensional Jordanian block is obvious:
\be
[L_{n},\Phi^{i}]=\left[z^{n+1}\partial_{z}+(n+1)z^{n}\left(\Delta+
\frac{\Lambda}{3}\right)\right]\Phi^{i}+(n+1) z^{n}\Phi^{i-1},
\ee
and
\be
[G_{r},\Phi^{(i)}]=[z^{r+1/3}(\delta_{\theta}-q^{2}\theta^{2}\partial_{z})-
(3r+1)q^{2}z^{r-2/3}\Delta\theta^{2}]\Phi^{i}-q^{2}\theta^{2}(3r+1)z^{r-2/3}
\Phi^{(i-1)}.
\ee
Here $1\leq i\leq m-1$, and the first member of the block, $\Phi^{(0)}$, is
a superprimary field. It is easy to show that (27) and (28) are satisfied
through the formal relation
\be
\Phi^{(i)}=\frac{1}{i!}\frac{d^{i}\Phi^{(0)}}{d\Delta^{i}}.
\ee
\section{Two point functions of Jordanian blocks}
Consider two Jordanian blocks of chiral quasi-primary fields
$\Phi_{1}$ and $\Phi_{2}$, with the weights $\Delta_{1}$ and
$\Delta_{2}$ and dimensions $p$ and $q$, respectively. As the
only closed subalgebra of the super Virasoro algebra the central extension
of which is trivial is formed by $[L_{-1}, L_{0}, G_{-1/3}]$, the correlator
of fields with different weights may be nonzero. According to [18],
\be
<\varphi_k^(0)\varphi^{\prime (0)}_{k'}>=a_K\frac{A_{kk'}(\Delta+\Delta'
+(k+k'-3)/3)^{B_{kk'}}}
{(z-z')^{\Delta+\Delta'+(k+k')/3}}=:a_Kf_{k,k'}(z-z'),
\ee
where $A_{kk'}$ and $B_{kk'}$ are the components of the following matrices:
\be
B=\pmatrix{0&0&0 \cr 0&0&1 \cr 0&1&1 \cr },\hspace{1 cm}
A=\pmatrix{1&1&1 \cr -1&1&1 \cr
q^{2}&-q^{2}&q^{2} \cr},
\ee
$a_{0}$, $a_{1}$, and $a_{2}$, are arbitrary constants, and
\be
K=k+k'\;\hbox{mod}\; 3.
\ee
The general form
of the two point functions of Jordanian blocks is then readily obtained,
using (29):
\be
<\varphi^{(i)}_{k}\varphi'^{(j)}_{k'}>=\frac{1}{i!}\frac{1}{j!}\frac{d^{i}}
{d\Delta^{i}}\frac{d^{j}}{d\Delta'^{j}}\frac{a_KA_{kk'}
(\Delta+\Delta'+(k+k'-3)/3)^{B_{kk'}}}{(z-z')^{\Delta+\Delta'+(k+k')/3}}.
\ee
Here $0\leq i\leq p-1$, and $0\leq j\leq q-1$. In this formal
differentiation, one should treat the constants $a_i$ as functions of
$\Delta$ and $\Delta'$. So, there will be other arbitrary constants
\be
a_i^{(j),(k)}:={{d^j}\over{d\Delta^j}}{{d^k}\over{d\Delta^k}}a_i
\ee
in these correlators.

To consider the correlators of the full field, one begins with
\be
<\varphi^{(00)}_{k\bar{k}}(z,\bar{z})\varphi^{\prime (00)}_{k'\bar{k}'}
(z,\bar{z})>=a_{K\bar{K}}q^{-k\bar{k}}
f_{k,k'}(z-z')\bar{f}_{\bar{k},\bar{k}'}(\bar{z}-\bar{z}'),
\ee
obtained in [18]. Here $f_{k,k'}(z-z')$ is defind in (30) and
$\bar{f}_{\bar{k},\bar{k}'}(\bar{z}-\bar{z}')$ is the same as
this with $\Delta\rightarrow\bar\Delta$ and
$\Delta'\rightarrow\bar{\Delta}'$. Also,
\bea K&=&k+k'\;\hbox{mod}\; 3\cr
 \bar{K}&=&\bar{k}+\bar{k}'\;\hbox{mod}\; 3.
\eea
Using the obvious generalization of (29), it is easy to see that
\be
<\varphi^{(ij)}_{k,\bar{k}}\varphi^{(lm)}_{k',\bar{k}'}>=\frac{1}{i!j!l!m!}
\frac{d^{i}}{d\Delta^{i}}\frac{d^{j}}{d\bar{\Delta}^{j}}\frac{d^{l}}
{d\Delta'^{l}}\frac{d^{m}}{d\bar{\Delta}'^{m}}[a_{K\bar K}f_{k,k'}
\bar f{\bar k,\bar k'}].
\ee
Again, one should treat $a_{K\bar K}$'s as formal functions of the weights,
so that differentiating them with respect to the weights introduces new
arbitrary parameters.

\end{document}